\documentclass
[twocolumn,amsmath,amssymb,amsfonts,prl,nopreprintnumbers]{revtex4}%
\usepackage[dvips]{graphicx}
\usepackage{dcolumn}
\usepackage{bm}
\usepackage{amsmath}
\usepackage{amsfonts}
\usepackage{amssymb}%
\preprint{Phys. Rev. Lett. {\bf 96}, 056402 (2006)}
\begin{document}
\title{Mapping of the Anisotropic Two-channel Anderson Model\qquad\qquad\qquad
\qquad\qquad\qquad\ onto a Fermi-Majorana bi-Resonant Level Model}
\author{C.~J.~Bolech}
\author{An\'{\i}bal Iucci}
\affiliation{Universit\'{e} de Gen\`{e}ve, DPMC, Quai Ernest Ansermet 24, CH-1211
Gen\`{e}ve 4, Suisse}
\date{April 4$^{\text{th}}$, 2005}

\begin{abstract}
We establish the correspondence between an extended version of the two-channel
Anderson model and a particular type of bi-resonant level model. For certain
values of the parameters the new model becomes quadratic. We calculate in
closed form the entropy and impurity occupation as functions of temperature
and identify the different physical energy scales of the problem. We show how,
as the temperature goes to zero, the model approaches a universal line of
fixed points non-Fermi liquid in nature.

\end{abstract}

\pacs{}
\maketitle

The challenge of single impurity models has a long history that started some
four decades ago with the study of the single-channel Kondo model
\cite{kondo1964}. The strongly-coupled physics of its low temperature regime
was determinant in the need to employ non-perturbative techniques to unravel
the physics and the nature of the infrared fixed point. The landmark
achievements in this respect were the numerical renormalization group study of
Wilson \cite{wilson1975} and the Bethe ansatz solution of Andrei and Wiegmann
\cite{andrei1980,wiegmann1980} on one hand, and the identification by Toulouse
\cite{toulouse1969,schlottmann1978} of a solvable point of the anisotropic
version of the model on the other hand. These three techniques are special,
because they allow the study of the crossover regime leading to the
unambiguous identification of the strongly-coupled Fermi-liquid fixed point of
the model. These tools were subsequently successfully applied to more general
models incorporating valence fluctuations \cite{Hewson}, or multiple channels
and non-Fermi-liquid characteristics \cite{cox1998}; the next logical step was
to consider models incorporating both elements simultaneously in order to
understand their interplay.

The ideas behind the two-channel Anderson model were first introduced in an
attempt to model the non-Fermi-liquid physics of certain \textrm{U}-based
heavy fermions like the \textrm{UBe}$_{13}$ compound \cite{cox1987}. Its
relationship to the two-channel Kondo model \cite{Nozieres1980} is as in the
case of the respective single-channel models: it not only captures the local
moment physics and provides a physical mechanism for moment formation, but at
the same time describes also other regimes in which mixed valence prevails all
the way down to the lowest temperatures. This is of great relevance because a
large number of compounds are believed to be near mixed valence and therefore
a good understanding of the \textit{full} Anderson model physics should prove
instrumental in the description of their phenomenology \cite{cox1998}. Whereas
the outermost $f$-shell of, for instance, \textrm{Ce} ions in typical
heavy-Fermion compounds is usually singly occupied, that of \textrm{U} ions is
believed to fluctuate between the $5f^{2}$ and $5f^{3}$ valence states. A
minimal model that takes into account spin-orbit and crystal-field effects
leads to modeling those two states with $\Gamma_{3}$ (flavor) and $\Gamma_{6}$
(spin) doublets that hybridize with $\Gamma_{8}$ conduction electrons
\cite{cox1998}.

On the other hand, the quest to a better understanding of non-Fermi-liquid
physics has recently permeated into the field of mesoscopics and there are
several attempts at realizing two-channel Kondo physics in the controlled and
highly-tunable realm of quantum dots \cite{berman1999,oreg2003,shah2003}.
Realizations based on the two-channel Anderson model may allow for a more
robust description of certain aspects of the physics of such devices, like the
charge fluctuations behind their capacitance lineshapes \cite{bolech2005b}.

Among the different non-perturbative techniques mentioned above, bosonization
--or Coulomb gas-- based mappings occupy a singular place \cite{Giamarchi}.
Their appeal is due to the elegance of the solution and the simplicity of the
picture that emerges from them; these qualities are invaluable in providing an
intuitive understanding of the physics and render them complementary to more
involved techniques like Bethe ansatz \cite{afl1983} or numerical
renormalization group \cite{wilson1975}. In this letter, we present a mapping
between the anisotropic two-channel Anderson impurity model and a
resonant-level Hamiltonian that for particular values of the parameters
becomes non-interacting. This property is analogous to the so-called Toulouse
point of the single-channel Kondo problem \cite{toulouse1969,schlottmann1978},
but displays characteristics of non-Fermi-liquid physics like in the
Emery-Kivelson mapping for the two-channel Kondo problem
\cite{emery1992,fabrizio1995,schofield1997,ye1997,vondelft1998}. Moreover, our
mapping of the two-channel Anderson model fully captures also the physics of
mixed valence, something achieved previously only in the infinite-$U$
single-channel case \cite{kotliar1996}. We confirm, in a very compact unified
language, all the predictions made recently for the model using a variety of
other non-perturbative techniques
\cite{schiller1998,bolech2002,johannesson2003,anders2005,jba2005,bolech2005a}.
We identify the two crossover energy scales of the model, below which the
physics is governed by a line of non-Fermi-liquid fixed points. We are able to
calculate both dynamical and thermodynamical quantities of interest over the
full temperature range, across both crossovers, and connect explicitly all
different high temperature regimes with the zero-temperature line of fixed points.

We are thus lead to consider the following generalization of the two-channel
Anderson Hamiltonian, $H=H_{\text{host}}+H_{\text{i+h}}+H_{3}\,$. The first
two terms correspond: to the band electrons ($\psi_{\alpha\sigma}^{\dagger}$)
the first one,%
\begin{equation}
H_{\text{host}}=\sum_{\alpha\sigma}\int dx~\psi_{\alpha\sigma}^{\dagger
}\left(  x\right)  \left(  -i\partial_{x}\right)  \psi_{\alpha\sigma}\left(
x\right)  ~\text{,}%
\end{equation}
and to the impurity ($X_{ab}$) the second one,%
\begin{multline}
H_{\text{i+h}}=\varepsilon_{s}\sum_{\sigma}X_{\sigma\sigma}+\varepsilon
_{f}\sum_{\alpha}X_{\bar{\alpha}\bar{\alpha}}+\\
+V\sum_{\alpha\sigma}\left[  X_{\sigma\bar{\alpha}}\psi_{\alpha\sigma}\left(
0\right)  +\psi_{\alpha\sigma}^{\dagger}\left(  0\right)  X_{\bar{\alpha
}\sigma}\right]  ~\text{.}%
\end{multline}
Taken toghether they constitute the standard two-channel model. Here we have
described the Hilbert space of the impurity using Hubbard-operator notation,
$X_{ab}=\left\vert a\right\rangle \left\langle b\right\vert $, where
$a,b=\left(  \sigma=\uparrow,\downarrow\right)  ,\left(  \bar{\alpha}=\bar
{+},\bar{-}\right)  $. The third term, $H_{3}=\sum_{\nu}H_{3}^{\nu}$,
involving density-density interactions in the different sectors ($\nu$), is
mainly added to break the two $SU\left(  2\right)  $ symmetries in spin and
flavor, introducing anisotropy as is standard in RG and Coulomb-gas analysis.
The impurity densities involved are%
\begin{gather}
X_{c}=X_{s\!f}=\sum_{\sigma}X_{\sigma\sigma}-\sum_{\alpha}X_{\bar{\alpha}%
\bar{\alpha}}~\text{,}\\
X_{s}=\sum_{\sigma}\sigma X_{\sigma\sigma}~\text{,\qquad\qquad\qquad}%
X_{f}=\sum_{\alpha}\alpha X_{\bar{\alpha}\bar{\alpha}}~\text{.}\nonumber
\end{gather}
The corresponding terms in the charge ($c$), spin ($s$), flavor ($f$), and
spin-flavor ($s\!f$) sectors are%
\begin{equation}
H_{3}^{\nu}=J_{\nu}^{3}X_{\nu}\sum_{\alpha\sigma\alpha^{\prime}\sigma^{\prime
}}\psi_{\alpha\sigma}^{\dagger}\left(  0\right)  \Upsilon_{\alpha\sigma
,\alpha^{\prime}\sigma^{\prime}}^{\nu}\psi_{\alpha^{\prime}\sigma^{\prime}%
}\left(  0\right)  ~\text{,}%
\end{equation}
where
\begin{equation}%
\begin{array}
[c]{lll}%
\Upsilon_{\alpha\sigma,\alpha^{\prime}\sigma^{\prime}}^{c}=\delta
_{\alpha\alpha^{\prime}}\delta_{\sigma\sigma^{\prime}}~\text{,} & \qquad &
\Upsilon_{\alpha\sigma,\alpha^{\prime}\sigma^{\prime}}^{s}=\delta
_{\alpha\alpha^{\prime}}\tau^{3}{}_{\sigma\sigma^{\prime}}~\text{,}\\
\Upsilon_{\alpha\sigma,\alpha^{\prime}\sigma^{\prime}}^{f}=\tau^{3}{}%
_{\alpha\alpha^{\prime}}\delta_{\sigma\sigma^{\prime}}~\text{,} & \qquad &
\Upsilon_{\alpha\sigma,\alpha^{\prime}\sigma^{\prime}}^{s\!f}=\tau^{3}%
{}_{\alpha\alpha^{\prime}}\tau^{3}{}_{\sigma\sigma^{\prime}}~\text{,}%
\end{array}
\end{equation}
and $\tau^{3}$ is the third Pauli matrix.

Many times, the physics of (1+1)-dimensional models is more transparent in a
bosonic representation \cite{Giamarchi}. The bosonization prescription reads
$\psi_{\alpha\sigma}\left(  x\right)  \approx e^{-i\phi_{\alpha\sigma}\left(
x\right)  }/\sqrt{2\pi a}$, with $a$ a regulator that goes to zero in the
continuum limit \footnote{More formally, one should include the so-called
Klein factors, but here we omit them for the sake of clarity of presentation.
If they were included, the refermionization would become much more involved
but all the final results would remain unchanged; a presentation of these
details will be provided in a forthcoming publication.}. It is convenient to
change basis in the bosonic fields according to $\phi_{\alpha\sigma}=(\phi
_{c}+\sigma\phi_{s}+\alpha\phi_{f}+\alpha\sigma\phi_{s\!f})/2$. Even more, we
find that remarkable simplifications are achieved by performing the canonical
transformation $U=\prod\nolimits_{\nu}e^{i\gamma_{\nu}\phi_{\nu}\left(
0\right)  X_{\nu}}$, with $2\gamma_{c}=\gamma_{s}=-\gamma_{f}=2\gamma
_{s\!f}=1/2$. This unitary transformation is a generalization of the one used
in the study of Kondo-type impurity exchange models. By choosing $J_{\nu}%
^{3}=\pi\gamma_{\nu}$, the first and third terms of the Hamiltonian yield%
\[
U\left(  H_{\text{host}}+H_{3}\right)  U^{\dagger}=\sum_{\nu}H_{0}^{\nu}%
\equiv\frac{1}{4\pi}\sum_{\nu}\int dx~\left(  \partial_{x}\phi_{\nu}\left(
x\right)  \right)  ^{2}%
\]
On the other hand, the impurity terms become%
\begin{multline*}
UH_{\text{i+h}}U^{\dagger}=\varepsilon_{s}\sum_{\sigma}X_{\sigma\sigma
}+\varepsilon_{f}\sum_{\alpha}X_{\bar{\alpha}\bar{\alpha}}+\\
+\frac{V}{\sqrt{2\pi a}}\left[  \left(  X_{\uparrow\bar{+}}+X_{\downarrow
\bar{-}}\right)  +e^{i\phi_{s\!f}}\left(  X_{\uparrow\bar{-}}+X_{\downarrow
\bar{+}}\right)  +\text{H.c.}\right]  ~\text{.}%
\end{multline*}

\begin{figure}
[t]
\begin{center}
\includegraphics[
height=0.6158in,
width=3.1661in
]%
{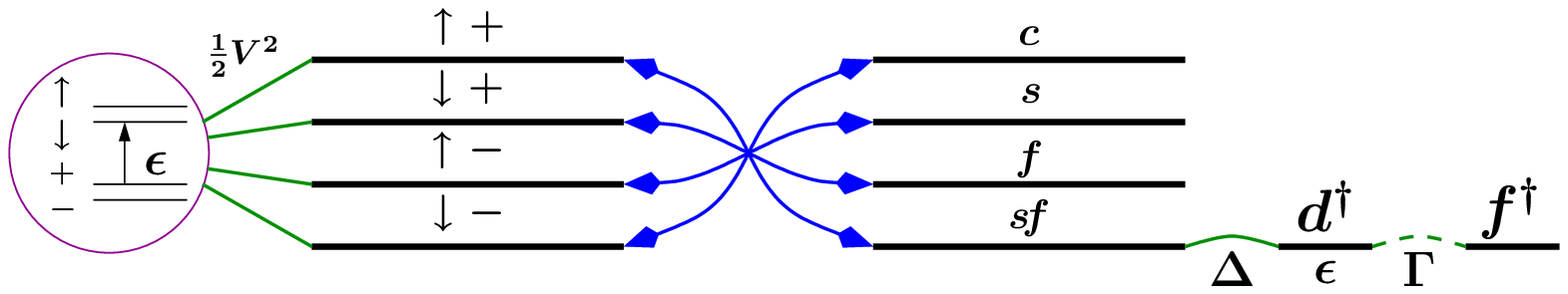}%
\caption{Pictorial representation of the anisotropic two-channel Anderson
impurity model before (right) and after (left) the mapping onto a
Fermi-Majorana bi-resonant level model.}%
\label{Fig:bires}%
\end{center}
\end{figure}

We can next refermionize the model introducing the fermionic operators
$f=X_{\bar{-}\bar{+}}+X_{\uparrow\downarrow}$ and $d=e^{-i\pi f^{\dagger}%
\!f}\left(  X_{\uparrow\bar{-}}+X_{\downarrow\bar{+}}\right)  $, plus the new
prescription $\psi_{\nu}\left(  x\right)  \approx e^{-i\pi\left(  f^{\dagger
}\!f+d^{\dagger}\!d\right)  }e^{-i\phi_{\nu}\left(  x\right)  }/\sqrt{2\pi a}%
$. We find that the impurity couples only to the spin-flavor sector and the
physics is governed by the following Fermi-Majorana bi-resonant level model
(see also Fig.~\ref{Fig:bires}):%
\begin{multline}
H_{\text{biRes}}=H_{0}^{s\!f}-\varepsilon~d^{\dagger}\!d+\varepsilon_{s}+\\
+\sqrt{2\Delta}\left[  \psi_{s\!f}^{\dagger}\left(  0\right)  d+d^{\dagger
}\psi_{s\!f}\left(  0\right)  \right]  +\\
+\sqrt{2\Gamma}\left(  f^{\dagger}-f\right)  \left(  d^{\dagger}+d\right)
~\text{,}%
\end{multline}
where $\varepsilon=\varepsilon_{s}-\varepsilon_{f}$, $\Delta=V^{2}/2$ and
$\Gamma=\Delta/2\pi a$. This is a purely quadratic model, on which
reintroducing the terms with non-zero $\lambda_{\nu}=\gamma_{\nu}-J_{\nu}%
^{3}/\pi$ would parametrize the deviations from the solvable point (cf. Refs.~%
[\onlinecite{clarke1993,sengupta1994}]%
). It is interesting to point out the similarities and differences between
this model and the Majorana resonant-level model that corresponds to the
solvable point of the two-channel Kondo model \cite{emery1992}. In both cases
the impurity hybridizes only with $\psi_{s\!f}^{\dagger}$, but for the
Anderson case the situation is more complex: two fermionic degrees of freedom
are required to represent the impurity. One ($d^{\dagger}$) with a chemical
potential $\varepsilon$ that vanishes at the intrinsic mixed-valence point and
goes off-resonance in the local-moment regimes. A second one ($f^{\dagger}$)
--related to the spin and flavor fluctuations-- that couples only via one of
its Majorana components and is always resonant in the absence of external
fields. As in the Kondo case, the other Majorana component of $f^{\dagger}$
exists completely decoupled from the rest of the system and will be
responsible for the \textit{fractional} residual impurity entropy that we
discuss below.%

It is a relatively simple task to extract the impurity thermodynamics and
correlation functions at the solvable point. The impurity free energy can be
conveniently computed using Pauli's trick of integrating over the coupling
constants. After some algebra, one arrives at%
\begin{align}
\Omega-\Omega_{0}  &  =\int_{0}^{1}\frac{d\lambda}{\lambda}~\left\langle
\lambda\left(  H_{\text{biRes}}-H_{0}^{s\!f}\right)  \right\rangle _{\lambda
}\nonumber\\
&  =-\int_{0}^{1}d\lambda~\frac{1}{\beta}\sum_{n\geq0}\frac{\partial_{\lambda
}D\left(  \omega_{n},\lambda\right)  }{D\left(  \omega_{n},\lambda\right)
}~\text{,}%
\end{align}
where%
\[
D\!\left(  \omega,\lambda\right)  =\lambda^{4}\!\left(  8\Gamma\Delta
+\Delta^{2}\omega\right)  +\lambda^{2}\!\left(  8\Gamma\omega+\varepsilon
^{2}\omega+2\Delta\omega^{2}\right)  +\omega^{3}%
\]
and $\omega_{n}=\pi(2n+1)/\beta$ are fermionic Matsubara frequencies.
Introducing a suitable regularization that can be removed later from the
actual physical quantities, one computes the different magnitudes of interest.
In particular, the impurity entropy is given by $S-S_{0}=\sum_{k}s\left(
z_{k}\right)  $ with%
\begin{equation}
s\left(  z\right)  =z\left[  \psi\left(  \frac{1}{2}+z\right)  -1\right]
-\ln\Gamma\left(  \frac{1}{2}+z\right)  +\frac{1}{2}\ln\pi
\end{equation}
and $\psi\left(  z\right)  $ the digamma function. Here $z_{k}=-\beta
\omega_{k}/2\pi$, with $\omega_{k=0,1,2}$ the three roots of $D\!\left(
\omega,1\right)  $. One finds that in the physical regime $\omega_{0}$ is real
while $\omega_{1,2}$ are complex conjugate of each other. That prompts us to
identify the Kondo and Schottky temperature scales:%
\begin{equation}
T_{\text{K}}\equiv-\omega_{0}/2\pi k_{\text{B}}\quad<\quad T_{\text{S}}%
\equiv\left\vert \omega_{1}\right\vert /2\pi k_{\text{B}}~\text{,}%
\end{equation}
so that%
\begin{equation}
S-S_{0}=s\left(  T_{\text{K}}/T\right)  +\left[  \text{Schottky contribution}%
\right]
\end{equation}
Remarkably, the function $s\left(  z\right)  $ is the same as that found for
the entropy of the two-channel Kondo model at the Emery-Kivelson point
\cite{fabrizio1995}. This is not completely unexpected, since Kondo is the low
energy effective theory for most part of the parameter regime of the
two-channel Anderson model.%
\begin{figure}
[t]
\begin{center}
\includegraphics[
height=2.3333in,
width=3.3278in
]%
{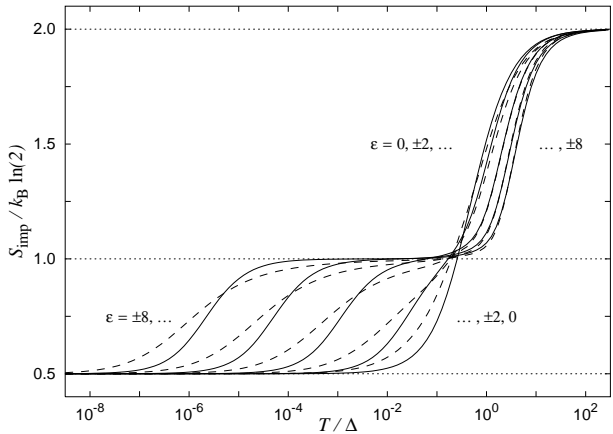}%
\caption{Impurity contribution to the entropy as a function of temperature for
different values of $\varepsilon$ ($0$, $\pm2$, $\pm4$, $\pm6$, $\pm8$). The
solid lines correspond to the results for the soluble point after
identification of the scales with the ones of the isotropic model. For the
sake of contrast, the corresponding curves for the latter are shown with
dashed lines for the same set of values of $\varepsilon$.}%
\label{Fig:entropy}%
\end{center}
\end{figure}
For the sake of illustration, in Fig.~\ref{Fig:entropy}, we show the impurity
contribution to the entropy as a function of temperature for the full range.
For the purpose of this figure, we have identified the two scales with those
of the isotropic model \cite{bolech2002}, which allows us to conveniently
display in the same plot the results for the latter. The figure illustrates
how, for large $\left\vert \varepsilon\right\vert $, the entropy of the
impurity is quenched in two stages as the temperature is lowered:
$k_{\text{B}}\ln4\rightarrow k_{\text{B}}\ln2\rightarrow k_{\text{B}}\ln
\sqrt{2}$. For small $\left\vert \varepsilon\right\vert $, the two quenching
steps coalesce on a single one. In all cases, the final value of the entropy
is the same, and indicative of the non-Fermi-liquid character of the
zero-temperature fixed points. Notice that the main difference between the
curves for the isotropic model and those for the anisotropic one at the
soluble point, reside in the shape of the second, or Kondo, quenching step.
This difference could be traced back to the absence of the leading irrelevant
operators at the soluble point (cf.~with the same situation in the case of the
two-channel Kondo model). In the language of specific heats, the $T\ln T$
leading terms are absent from the soluble anisotropic model and can be
recovered using perturbation theory in $\lambda_{\nu}$ (cf. Refs.~%
[\onlinecite{emery1992,sengupta1994}]%
). The same holds true for the leading logarithms in the magnetic and flavor
susceptibilities \cite{clarke1993}.%

\begin{figure}
[t]
\begin{center}
\includegraphics[
height=2.3328in,
width=3.3275in
]%
{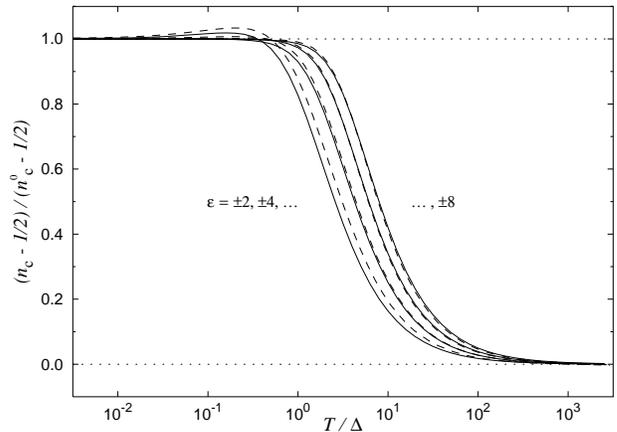}%
\caption{Normalized curves of the impurity charge valence as a function of
temperature. The results for the soluble anisotropic case (solid lines) and
Bethe ansatz results for the isotropic case (dashed lines) are displayed for
different values of $\varepsilon$ ($\pm2$, $\pm4$, $\pm6$, $\pm8$).}%
\label{Fig:valence}%
\end{center}
\end{figure}
Another quantity of interest is the impurity charge valence $n_{c}%
=\partial\Omega/\partial\varepsilon$ given by%
\begin{equation}
n_{c}-n_{c,0}=\frac{1}{\beta}\sum_{k}\psi\left(  \frac{1}{2}+z_{k}\right)
\partial_{\varepsilon}z_{k}~\text{.}%
\end{equation}
This is an aspect of the physics inherent to Anderson type models and of
particular relevance in their application in the context of quantum dots and
other mesoscopic systems that allow direct measurements of it
\cite{bolech2005b}. The valence starts at $n_{c,0}=1/2$ for high temperature
($T\gg T_{\text{S}}$) and evolves to reach finally a certain zero-temperature
value $n_{c}^{0}\equiv n_{c}\left(  \varepsilon\right)  _{T=0}$ that labels
the line of fixed points of the model. Figure~\ref{Fig:valence} shows
normalized curves for the temperature dependence of the impurity charge
valence. The quenching of the valence fluctuations takes place at the
characteristic scale $T_{\text{S}}$. The correspondence between the
bosonization results and the results for the isotropic case is rather good for
$\left\vert \varepsilon\right\vert $ large, and the differences for small
$\left\vert \varepsilon\right\vert $ are in part due to the difficulty for
identifying the scales of the two models. Nevertheless, subtle aspects of the
small $\left\vert \varepsilon\right\vert $ curves, like the `overshoot' of the
curves at intermediate temperatures $T\lesssim T_{\text{S}}$, are also present
at the soluble point. All this shows that the isotropic model and the soluble
anisotropic point not only share the same infrared fixed-point behavior,
albeit with different irrelevant operators content, but also display matchable
generic ultraviolet physics.

In summary, we have shown that the anisotropic two-channel Anderson model can
be solved exactly for particular values of the coupling constants in the
\textit{extra} $H_{3}$ term of the Hamiltonian. Using bosonization, we
demonstrated that the problem reduces to the study of a non-interacting
Fermi-Majorana bi-resonant level model. Deviations from the solvable point can
be taken into account using perturbation theory in, call it, $\delta H_{3}$.
The advantage of the method is evident: closed analytical expressions can be
derived for the full temperature crossovers of the different quantities of
interest; this sets the approach apart from other non-perturbative techniques
applied to the model in the past. Although the fixed point line of the
solvable anisotropic case is the same as for the usual isotropic model
(\textit{i.e.}, anisotropy is irrelevant), the leading irrelevant operator
content of the anisotropic model is more restricted, --which lies behind its
greater simplicity. A manifestation of this difference is found, for instance,
in the results we presented for the impurity entropy. On a different front,
and as compared with pure-exchange type of models like the two-channel Kondo,
the Anderson model brings in as well the physics of mixed valence and charge
fluctuations. We have shown that the essential aspects of this physics are
again well captured by the bi-resonant level Hamiltonian. In a future
contribution, we plan to give more specialized technical details of the
Abelian bosonization procedure and discuss the different field
susceptibilities \footnote{In the presence of fields, the degree of $D\left(
\omega,1\right)  $ increases by one and there is a fourth root that appears
and sets the temperature scale below which the system is driven away from the
non-Fermi liquid line of fixed points. Notice that the fields introduce a link
to the otherwise decoupled Majorana component of $f^{\dagger}$.}. This work
opens up multiple other possibilities for the study of two-channel Anderson
models as applied to the physics of heavy-fermions and mesoscopic systems. One
may, for instance, consider the behavior of more than one impurity and, in
particular, the case of two-channel Anderson lattices.

\begin{acknowledgments}
This work was partly supported by the Swiss National Science Foundation under
MaNEP and Division II.
\end{acknowledgments}


\end{document}